\documentclass[a4paper,12pt, epsfig]{article}
\usepackage{epsfig}
\usepackage{amssymb,latexsym}
\usepackage{amsfonts}
\usepackage{amsmath}
\usepackage{epstopdf}

\newskip\humongous \humongous=0pt plus 1000pt minus 1000pt

\newif\ifdtup

\catcode`\@=11

\@addtoreset{equation}{section}
\def\theequation{\thesection.\arabic{equation}}

\def\@normalsize{\@setsize\normalsize{15pt}\xiipt\@xiipt
\abovedisplayskip 14pt plus3pt minus3pt%
\belowdisplayskip \abovedisplayskip
\abovedisplayshortskip \z@ plus3pt%
\belowdisplayshortskip 7pt plus3.5pt minus0pt}

\def\small{\@setsize\small{13.6pt}\xipt\@xipt
\abovedisplayskip 13pt plus3pt minus3pt%
\belowdisplayskip \abovedisplayskip
\abovedisplayshortskip \z@ plus3pt%
\belowdisplayshortskip 7pt plus3.5pt minus0pt
\def\@listi{\parsep 4.5pt plus 2pt minus 1pt
      \itemsep \parsep
      \topsep 9pt plus 3pt minus 3pt}}

\relax

\catcode`@=12

\catcode`\@=11

\def\section{\@startsection{section}{1}{\z@}{3.5ex plus 1ex minus
    .2ex}{2.3ex plus .2ex}{\large\bf}}

\def\thesection{\arabic{section}}
\def\thesubsection{\arabic{section}.\arabic{subsection}}

\def\appendix{\setcounter{section}{0}
  \def\thesection{Appendix \Alph{section}}
  \def\thesubsection{\Alph{section}.\arabic{subsection}}
  \def\theequation{\Alph{section}.\arabic{equation}}}

\def\SymBoxes#1#2#3#4{\newdimen\un@t \un@t#3%
\raisebox{#1}{\rule{#2\un@t}{#4}\hskip-#2\un@t
\@tempdimb\un@t \advance\@tempdimb by-#4\@tempcntb#2\relax%
\@whilenum{\@tempcntb>0}\do{
\rule{#4}{\un@t}\hskip\@tempdimb \advance\@tempcntb by\m@ne}%
\hskip-#2\un@t \rule[\un@t]{#2\un@t}{#4}%
\rule[\un@t]{#4}{#4}\hskip-#4
\rule{#4}{\un@t}}\hskip-#4}                

\begin{document}


\newcommand{\dd}{\textrm{d}}

\newcommand{\beq}{\begin{equation}}
\newcommand{\eeq}{\end{equation}}
\newcommand{\bea}{\begin{eqnarray}}
\newcommand{\eea}{\end{eqnarray}}
\newcommand{\beas}{\begin{eqnarray*}}
\newcommand{\eeas}{\end{eqnarray*}}
\newcommand{\defi}{\stackrel{\rm def}{=}}
\newcommand{\non}{\nonumber}
\newcommand{\bquo}{\begin{quote}}
\newcommand{\enqu}{\end{quote}}
\newcommand{\tc}[1]{\textcolor{blue}{#1}}
\renewcommand{\(}{\begin{equation}}
\renewcommand{\)}{\end{equation}}
\def\de{\partial}
\def\Om{\ensuremath{\Omega}}
\def\Tr{ \hbox{\rm Tr}}
\def\rc{ \hbox{$r_{\rm c}$}}
\def\H{ \hbox{\rm H}}
\def\HE{ \hbox{$\rm H^{even}$}}
\def\HO{ \hbox{$\rm H^{odd}$}}
\def\HEO{ \hbox{$\rm H^{even/odd}$}}
\def\HOE{ \hbox{$\rm H^{odd/even}$}}
\def\HHO{ \hbox{$\rm H_H^{odd}$}}
\def\HHEO{ \hbox{$\rm H_H^{even/odd}$}}
\def\HHOE{ \hbox{$\rm H_H^{odd/even}$}}
\def\K{ \hbox{\rm K}}
\def\Im{ \hbox{\rm Im}}
\def\Ker{ \hbox{\rm Ker}}
\def\const{\hbox {\rm const.}}
\def\o{\over}
\def\im{\hbox{\rm Im}}
\def\re{\hbox{\rm Re}}
\def\bra{\langle}\def\ket{\rangle}
\def\Arg{\hbox {\rm Arg}}
\def\exo{\hbox {\rm exp}}
\def\diag{\hbox{\rm diag}}
\def\longvert{{\rule[-2mm]{0.1mm}{7mm}}\,}
\def\a{\alpha}
\def\b{\beta}
\def\e{\epsilon}
\def\l{\lambda}
\def\ol{{\overline{\lambda}}}
\def\ochi{{\overline{\chi}}}
\def\th{\theta}
\def\s{\sigma}
\def\oth{\overline{\theta}}
\def\ad{{\dot{\alpha}}}
\def\bd{{\dot{\beta}}}
\def\oD{\overline{D}}
\def\opsi{\overline{\psi}}
\def\dag{{}^{\dagger}}
\def\tq{{\widetilde q}}
\def\L{{\mathcal{L}}}
\def\p{{}^{\prime}}
\def\W{W}
\def\N{{\cal N}}
\def\hsp{,\hspace{.7cm}}
\def\bo{\ensuremath{\hat{b}_1}}
\def\bfo{\ensuremath{\hat{b}_4}}
\def\co{\ensuremath{\hat{c}_1}}
\def\cfo{\ensuremath{\hat{c}_4}}
\newcommand{\C}{\ensuremath{\mathbb C}}
\newcommand{\Z}{\ensuremath{\mathbb Z}}
\newcommand{\R}{\ensuremath{\mathbb R}}
\newcommand{\rp}{\ensuremath{\mathbb {RP}}}
\newcommand{\cp}{\ensuremath{\mathbb {CP}}}
\newcommand{\vac}{\ensuremath{|0\rangle}}
\newcommand{\vact}{\ensuremath{|00\rangle}                    }
\newcommand{\oc}{\ensuremath{\overline{c}}}
\renewcommand{\cos}{\textrm{cos}}
\renewcommand{\sin}{\textrm{sin}}

\newcommand{\Vol}{\textrm{Vol}}

\newcommand{\half}{\frac{1}{2}}

\def\changed#1{{\bf #1}}

\begin{titlepage}

\def\thefootnote{\fnsymbol{footnote}}

\begin{center}
{\large {\bf
The Reactor Anomaly after Daya Bay and RENO
  } }

\bigskip

\bigskip

{\large \noindent Emilio
Ciuffoli$^{1}$\footnote{ciuffoli@ihep.ac.cn}, Jarah
Evslin$^{1}$\footnote{\texttt{jarah@ihep.ac.cn}}, Hong Li$^{2,
3}$\footnote{\texttt{hongli@ihep.ac.cn}}}
\end{center}

\renewcommand{\thefootnote}{\arabic{footnote}}

\vskip.7cm

\begin{center}
\vspace{0em} {\small \em  {1) TPCSF, IHEP, Chinese Academy of Sciences\\
YuQuan Lu 19(B), Beijing 100049, P.R.China

\vspace{.2cm}

2) Key Laboratory of Particle Astrophysics,
IHEP, \\ Chinese Academy of Sciences,
P.O.Box 918-3, Beijing 100049, P.R.China

\vspace{.2cm}

3) National Astronomical
Observatories, Chinese Academy of Sciences,\\
 Beijing 100012, P.R.China

\vspace{.2cm}

}}

\vskip .4cm

\end{center}

\vspace{1.0cm}

\noindent
\begin{center} {\bf Abstract} \end{center}

\noindent Gallium and short baseline reactor neutrino experiments indicate a short-distance anomalous disappearance of electron antineutrinos  which, if interpreted in terms of neutrino oscillations, would lead to a sterile neutrino mass inconsistent with standard cosmological models.  This anomaly is difficult to measure at 1 km baseline experiments because its disappearance effects are degenerate with that of $\theta_{13}$.  The flux normalization independent measurement of $\theta_{13}$ at Daya Bay breaks this degeneracy, allowing an unambiguous differentiation of 1-3 neutrino oscillations and the anomalous disappearance at Double Chooz and RENO.  The resulting anomaly is consistent with that found at very short baselines and suggests a downward revision of RENO's result for $\theta_{13}$.  A MCMC global analysis of current cosmological data shows that a quintom cosmology is just compatible at $2\sigma$ with a sterile neutrino with the right mass to reproduce the reactor anomaly and to a lesser extent the gallium and LSND/MiniBooNE anomalies. However models in which the sterile neutrino acquires a chameleon mass easily satisfy the cosmological bounds and also reduce the tension between LSND and KARMEN.

\vfill

\begin{flushleft}
{\today}
\end{flushleft}
\end{titlepage}

\hfill{}


\setcounter{footnote}{0}

\section{Introduction}
During the past 20 years, the gallium experiments GALLEX \cite{gallex} and SAGE \cite{sage} have observed\footnote{Here we use the latest calibrations \cite{gallexcalib}.} 14$\pm$5\% \cite{gallium} less neutrinos than expected from radioactive sources inside of the detector.  While Lorentz violating \cite{minosv}, CP violating \cite{lsnd,minibooneanom} and CPT violating \cite{minosanom} neutrino anomalies have disappeared in the last few months \cite{zichichi,miniboonetuttobene,minostuttobene}, the gallium disappearance anomaly has been corroborated by a new calculation of reactor neutrino fluxes \cite{nuovoflusso}.  This new calculation suggests a 3.5\% increase in theoretical fluxes from nuclear reactors, leading to a total flux which is about 5.7$\pm$2.3\% higher than that observed at short baseline reactor experiments \cite{reattoreanom}.   We will refer to this missing reactor neutrino flux as the reactor anomaly.  The reactor and gallium anomalies stand out among neutrino anomalies because they occur in a relatively accessible environment\footnote{This is in contrast with the low energy solar neutrino deficit \cite{smirnov} and the horizontal flux excess \cite{icecube}.}, which can and will be directly probed by future experiments, and yet their simplest interpretation in terms of oscillations of a 1 or more eV sterile neutrino \cite{giuntireview} is in conflict with standard cosmology (for a recent analysis see Ref.~\cite{sterilecosm}).

Not all reactor neutrino disappearance is anomalous, reactor neutrinos also disappear as a result of oscillation between the three standard neutrino flavors.  At distances below about 100 meters, all such oscillation is well within the experimental errors of current measurements.  Above 100 meters 1-3 neutrino oscillation is relevant and above about 2 kilometers 1-2 oscillation is also significant.  As the 1-3 mixing angle $\theta_{13}$ was poorly measured until quite recently, it has not been possible to disentangle the reactor anomaly from 1-3 oscillation in 1 km baseline reactor experiments.  This situation changed 3 months ago when the Daya Bay collaboration measured $\theta_{13}$ by comparing neutrino fluxes observed at identical detectors at various distances from an array of reactors \cite{dayabay}.   Following the proposal of Ref.~\cite{piureattori} this allowed them to determine $\theta_{13}$ independently of any assumptions about the overall flux normalization.    Recently, with 3 months more data, the Daya Bay collaboration has determined $\theta_{13}$ yet more precisely \cite{neut2012}.

In this note we will combine the results of the reactor neutrino experiments Double Chooz \cite{doublechooz}, Daya Bay \cite{dayabay,neut2012} and RENO \cite{reno,nuturn} to simultaneously estimate the reactor anomaly and $\theta_{13}$.   In particular the flux independent mixing angle obtained by Daya Bay allows for an unambiguous separation of standard mixing and anomalous disappearance at the other experiments.    As has already been seen in the global fits in Refs.~\cite{globale1,globale2}, greater reactor anomalies imply smaller mixing angles.  Using the reactor anomaly measured at short baseline experiments we will find that the preferred mixing angle is lower than RENO's result.  It is easy to extend our analysis to Palo Verde \cite{paloverde} and Chooz \cite{chooz} experiments and to gallium experiments; however, as we will see in  Fig.~\ref{tuttifig}, this extension does not change our results appreciably, so we will focus on the Daya Bay, RENO and Double Chooz results, confronting them with the reactor anomaly discussed in \cite{reattoreanom}.  We did not use the latest data release from Double Chooz \cite{doublechoozneut2012} which includes 142 days in which $\sin^2(2\theta_{13})$ nearly doubled with respect to the first 86 days and is highly dependent upon the analysis used.  

We will then consider theoretical models which may explain this anomaly consistently with cosmological constraints.  We will see that a quintom dark energy model with a massive sterile neutrino is consistent with these anomalies at about 2$\sigma$.  On the other hand a neutrino dark energy model \cite{neutdarke} in which a sterile neutrino mass is proportional to the density of its environment \cite{neal04} easily satisfies cosmological constraints and also reduces the tension between the positive appearance data at LSND and MiniBooNE and the negative results at KARMEN.


After this paper was completed we received the preprints \cite{tortola,foglinuovo} which combine data from the same three experiments.  Their analysis differs from ours in the treatments of the reactor anomaly and the total normalization of the theoretical reactor fluxes.   Our analysis is 2-dimensional, simultaneously fitting for the anomalous disappearance and 1-3 oscillation induced electron antineutrino disappearance.


\section{Disappearance Results from RENO}
The RENO experiment consists of two detectors, one near and one far, placed near an array of six nuclear reactors.  RENO's results \cite{reno} have been posted on the archive in two versions and have been extended at a recent talk \cite{nuturn} at the conference $\nu$TURN.  As they do not provide information about the reactor fluxes, or even about the calibration of the machines as in Ref.~\cite{dayafeb}, no single reference is sufficient to reproduce the claimed results.  However, by combining information from, for example, the most recent arXiv version and the $\nu$TURN talk we were able to reproduce the best mixing angle and flux deficit with an error of about one part in one thousand and to reproduce the $\chi^2$ value to within the widths of the lines in their graph.    

Following the analysis of Ref.~\cite{reno}, we have included all neutrinos with prompt energies up to 12 MeV, however it has been claimed in Ref.~\cite{thierry} that the deficit of neutrinos above 6 MeV is largely unrelated to neutrino oscillation and serves to artificially deflate the best fit mixing angle.  A adapting a 6 MeV cut would make the RENO result more compatible with the rate only analysis of the new Double Chooz data \cite{doublechoozneut2012} which we have not used, increasing our best fit values of both $\theta_{13}$ and the reactor anomaly.

Fig.~3 of Ref.~\cite{reno} provides the oscillation probability, for a given value of $\theta_{13}$, as a function of the weighted baseline $L_i$, which is the average distance traveled by a neutrino detected at the $i$th detector.
\ Ignoring $1-2$ neutrino oscillations, which is reasonable within the errors at this baseline, the survival probability of neutrinos arriving at the $i$th detector is
\beq
P_i=\sum_{j=1}^6 f_{ij} \left[ 1-\sin^2(2\theta_{13})\int \sin^2\left(\frac{\Delta m_{13}^2 d_{ij}}{4E}\right)\rho_{j}(E)dE\right] \label{sparisce}
\eeq
where $f_{ij}$ is the fraction of the neutrino flux observed at the $i$th detector which originated at the $j$th reactor, given in table 1 of version 1 of \cite{reno} and also at the talk \cite{nuturn}, $\delta m_{13}^2$ is the difference between the squared masses of the first and third neutrino flavors and $\rho_{j}(E)$ is the fractional energy distribution of neutrinos emitted from the $j$th reactor.  This is not equal to the quantity which is plotted in Fig.~3, which is the survival probability at the weighted baseline
\beq
P_i\neq 1-\sin^2(2\theta_{13})\int \sin^2\left(\frac{\Delta m_{13}^2 L_{i}}{4E}\right)\rho(E)dE.
\eeq
Nonetheless, as we will see the information contained in Fig.~3 is essential to reproduce RENO's analysis and to extrapolate the observed reactor anomaly.  


To determine which values of the anomalous neutrino deficit and $\theta_{13}$ best fit RENO's data, we used the formula (\ref{sparisce}).  We used the fractional fluxes reported in table 1 of version 1 of \cite{reno} or equivalently page 6 of \cite{nuturn} and the normalization of the expected fluxes from figure 3 of version 2.  There are several inequivalent ways of estimating the energy distribution $\rho_j(E)$, without making use of the unreleased reactor data.  One can add 780 keV to the prompt energy spectrum reported in Fig.~4
\beq
E_{\overline{\nu}}=E_{\rm{prompt}}+780{\rm{\ keV}}.
\eeq
However the statistical errors are large and the quality of the energy calibration is unknown.  We found that in analyzing both RENO and Daya Bay data our fits matched those of the experimental groups more closely when we used the sample neutrino energy spectra of Fig.~1.9 of Ref.~\cite{daya2007}.

To estimate errors we used the pull parameter method of the second version of Ref.~\cite{reno}, with $\chi^2$ given by their Eq.~(2).
However as we did not wish to impose any theoretical bias upon the overall reactor flux, we did not restrict its value.  We used the systematic errors reported in the second paper.   When we optimized the reactor flux within the region that they described, our resulting $\chi^2$ curve agreed with that shown on the top of RENO's Fig.~3 to within the widths of their lines.

The flux normalization and $\theta_{13}$ are two free parameters.  A flux only analysis, such as that performed by RENO, has only two data points, the fluxes observed at the two detectors.  Therefore the effective number of degrees of freedom is $2-2=0$, reflecting the familiar fact that with two parameters one can generically fit two unknowns.  Thus it is of no surprise that the minimum $\chi^2$ is equal to 0.  This is true in their case, in which the reactor flux is optimized and then $\theta_{13}$ is fit and it is true in our case, in which we wish to use RENO's data to fit both the flux deficit and $\theta_{13}$. 

 Using RENO's near and far detector fluxes reported in their Fig. 3 to determine the neutrino flux deficit and $\theta_{13}$ we obtain essentially the same fit as was obtained by the RENO collaboration, seen as a horizontal stripe in Fig.~\ref{noreno}.   The stripe is horizontal because, without knowledge of the theoretical flux, the reactor anomaly cannot be estimated.

However, RENO has tentatively reported the theoretical flux.  Most recently, in the talk~\cite{nuturn} they have provided a preliminary estimate of the anomalous flux deficit.  Using their value of $\theta_{13}$ they estimate this deficit to be 6\%.  
We have performed a 2-dimensional fit on their new data, providing confidence intervals for both the anomalous flux deficit and $\theta_{13}$.  Our result is the ellipse in the left panel of Fig.~\ref{tuttifig}.  The center of the ellipse indeed corresponds to the best fit found by the RENO collaboration.  Note that a greater anomaly means that less electron neutrinos are expected, because for example they have converted into sterile neutrinos, and so the deficit at both detectors is reduced.   This can be compensated at the far detector by decreasing $\theta_{13}$, which explains the inclination of the ellipse and was already evident in the global analyses of Refs. \cite{globale1,globale2}. 


\section{Disappearance Results from Double Chooz}

Although next year Double Chooz will have both a near and far detector, for now the data contains results from a single detector, which is compared with theoretical fluxes.   The Double Chooz collaboration analyzed their data in a number of different ways, arriving at different best fit mixing angles.  For the sake of uniformity, we will perform a pure rate analysis, using only the number of neutrinos observed, the theoretical flux and the baseline, even though this leads to large systematic errors \cite{doublechooz}.  As there is only one detector and both reactors are the same distance from that detector, in this case the spectrum dependence in the survival probability only enters via the energy-weighted quantity
\beq
\int \rho(E)\sin^2\left(\frac{\Delta m_{13}^2 L_{i}}{4E}\right)dE
\eeq
which is easily extracted from Ref.~\cite{doublechooz}.  Thus in this case we are not required to make any arbitrary assumptions about the isotope fractions in the reactors.

\begin{figure} 
\begin{center}
\includegraphics[scale=.42]{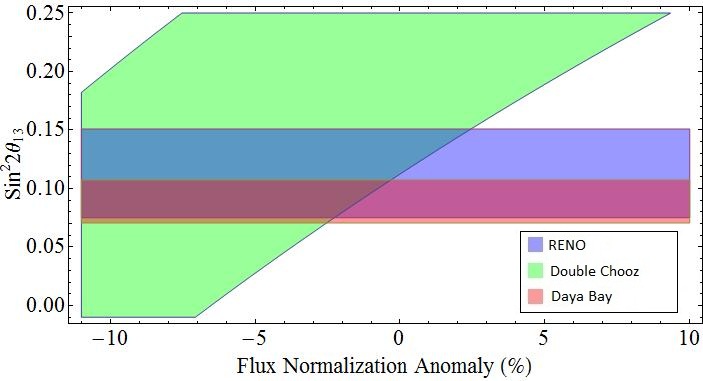}\hspace{0.1cm}\includegraphics[scale=.34]{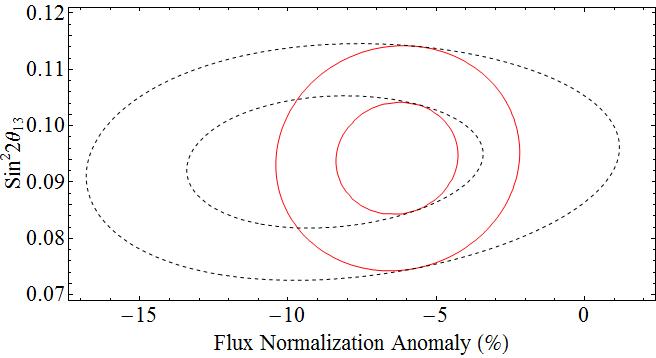}
\caption{We illustrate the values of the anomalous neutrino flux disappearance and $\sin^2(2\theta_{13})$ at Double Chooz, RENO and Daya Bay without using the absolute normalization of the reactor flux at RENO.  A $0\%$ anomaly corresponds to the preferred theoretical flux normalization.  The uncertainty in the theoretical normalization is not shown.  The 90\% confidence regions of the three experiments are shown in the left panel.  The right panel shows the 1 and 2$\sigma$ regions of the combined results of Daya Bay, Double Chooz, RENO (dotted curve) and also of these three experiments plus very short baseline reactor experiments (purple curve).}
\label{noreno}
\end{center}
\end{figure}

The Double Chooz paper Ref.~\cite{doublechooz} was written after the upward revision in theoretical fluxes in Ref.~\cite{nuovoflusso}, in fact the upward revision was motived by its use at Double Chooz.  Nonetheless Double Chooz did not normalize its data using the new fluxes, but instead using a hypothetical version of the Bugey4 experiment \cite{bugey4} with modified abundances for the four main fissioning isotopes to fix  the flux normalization by hand.  However they also reported the theoretical calculation for the flux using Ref.~\cite{nuovoflusso}, whose comparison with the observed flux we have used to determine their neutrino deficit.  

In the analysis  we use only a single data point, the total neutrino flux observed at the detector.  We fit this data point using 2 parameters, the anomalous flux deficit and $\theta_{13}$.  Therefore we have one constraint on two unknowns, and so there is a 1-dimensional curve of solutions with $\chi^2=0$.  To reproduce the value of $\theta_{13}$ reported by Double Chooz in Ref.~\cite{doublechooz} one would need to also impose an anomalous flux deficit equal to that measured by Bugey4, which according to \cite{reattoreanom} was about 6\% but according to \cite{doublechooz} for the isotope fractions at the Chooz reactors was about 8\%.  While Bugey4's flux measurement does represent a typical value of the reactor anomaly and while its errors are relatively small, we will not adopt this strategy.  The flux deficit will instead be left as a free parameter in all plots, which can then be compared with the reactor anomaly determined not only at Bugey4 but at all very short baseline reactor experiments and if desired gallium experiments and RENO.

\begin{figure} 
\begin{center}
\includegraphics[scale=.44]{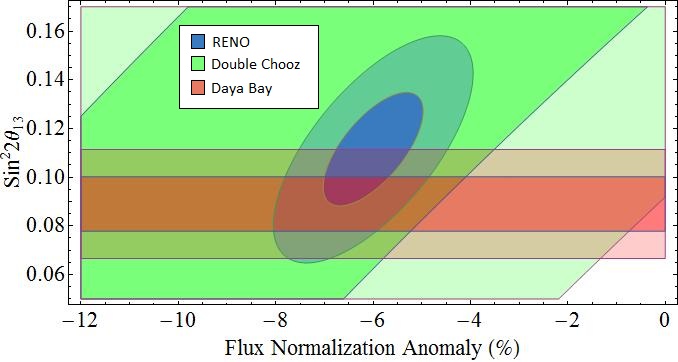} \hspace{0.1cm}\includegraphics[scale=.31]{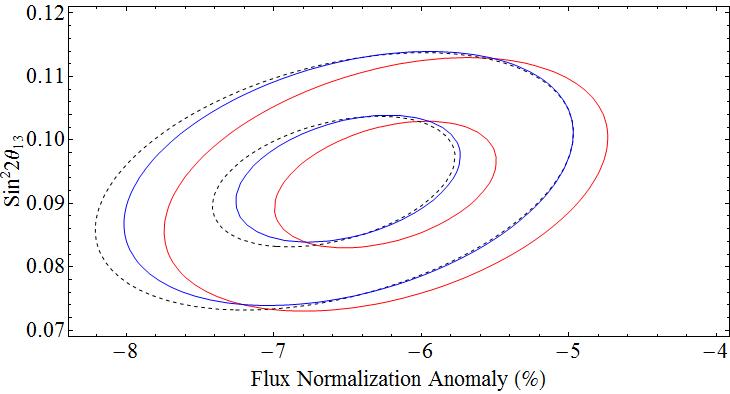}
\caption{We illustrate the values of the anomalous neutrino flux disappearance and $\sin^2(2\theta_{13})$ obtained using RENO's preliminary result for the flux deficit \cite{nuturn}.  Left Panel: The results of Double Chooz, RENO and Daya Bay experiments are shown (1 and 2$\sigma$ regions). Right panel: 1 and 2$\sigma$ regions of the combined results of only Daya Bay, Double Chooz and RENO (dotted curve),  of these plus the very short baseline experiments (red curve) and finally all of the above plus the Chooz, Palo Verde and gallium experiments (blue curve).}
\label{tuttifig}
\end{center}
\end{figure}


We have performed our analysis without using RENO's preliminary theoretical flux.  Our results are illustrated in Fig.~\ref{noreno}.  The right panel is a combined fit from the three experiments.  The corresponding ellipse is nearly horizontal, implying that the reactor anomaly and $\theta_{13}$ are nearly decoupled.  This decoupling is a result of the fact that the only information on the absolute normalization of the flux comes from Double Chooz, whose errors are quite large.  Therefore the preferred value of the reactor anomaly is about equal to that of Double Chooz, and the preferred value of $\theta_{13}$ is nearly a weighted average of that obtained at the three experiments.  If in this case one ignores the small correlation between the reactor anomaly and the mixing angle, then the analysis will be reasonably independent of the flux normalizations.  With this approximation we would reproduce the analysis of these three experiments performed in Ref.~\cite{tortola}.  

However if we include the preliminary flux normalization reported by the RENO collaboration in their talk \cite{nuturn} then we find by far the strongest evidence yet for a reactor anomaly, as seen in Fig.~\ref{tuttifig}.


One may try to incorporate the above 1 km baseline fits of the mixing angle $\theta_{13}$ and the reactor flux deficit with complimentary data from the flux normalization-independent analysis by Daya Bay \cite{dayabay}, the reactor anomaly at short baseline reactor experiments \cite{reattoreanom}, the neutrino deficit observed at gallium experiments \cite{gallium} and global fits of experimental data more than 10 months old \cite{globale1,globale2}.

The Daya Bay data is easily analyzed in this framework.  As it depends only upon the mixing angle $\theta_{13}$, the confidence intervals translate into a simple horizontal stripe on our Figs.~\ref{noreno} and \ref{tuttifig}.  Similarly the short distance reactor and gallium anomalies are independent of $\theta_{13}$, and so they correspond to vertical stripes.  The global analyses are somewhat more difficult to adapt to our setting, in which the reactor flux normalization is not fixed.  However Ref.~\cite{globale1} produces confidence intervals for $\theta_{13}$ for both the new reactor fluxes of Ref.~\cite{nuovoflusso} and the old normalization, which corresponds to a 3-3.5\% lower flux.   An adaptation of the global analysis of Ref.~\cite{globale2} to an arbitrary reactor anomaly is more difficult.  The mixing angle cited in their abstract includes a fit of short distance reactor data which effectively already changes the flux normalization.  Only the mixing angle which they quote in their Eq.~(3), in which short baseline reactor data is excluded, uses the flux normalization of Ref.~\cite{nuovoflusso}.  It is consistent with the analysis of Ref.~\cite{globale1} for this flux choice.

\begin{figure} 
\begin{center}
\includegraphics[scale=.61]{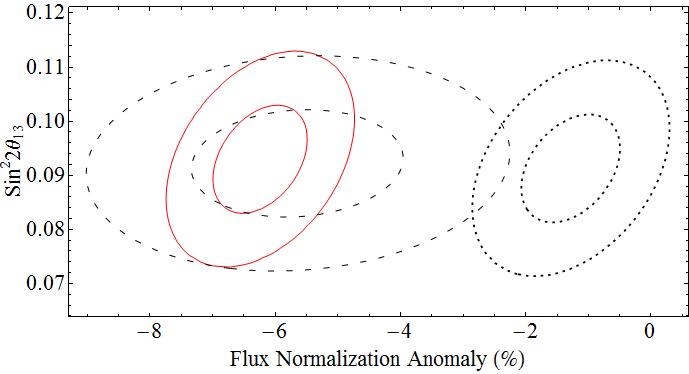}
\caption{We illustrate the 1 and 2$\sigma$ compatibility regions obtained by combining the results of Double Chooz, Daya Bay, RENO, Chooz, Palo Verde, very short baseline experiments and gallium experiments with (solid purple curve) and without (dashed blue curve) RENO's preliminary theoretical flux reported in the talk \cite{nuturn}.  To further illustrate the central role played by the absolute flux normalization at RENO in the determination of the flux anomaly we have included the black dotted curve, corresponding to a hypothetical scenario in which RENO reports a vanishing anomalous flux deficit.}
\label{piccolo}
\end{center}
\end{figure}


In Fig.~\ref{piccolo} we have plotted the 1 and 2$\sigma$ allowed regions obtained by combining the results of Double Chooz, Daya Bay, RENO, Chooz, Palo Verde, very short baseline experiments and gallium experiments with and without RENO's preliminary theoretical flux reported in the talk \cite{nuturn}.  This figure illustrates that kilometer baseline reactor experiments may provide the most convincing evidence yet for a reactor anomaly.  For now this result is highly dependent upon preliminary normalization results from RENO, a situation which will be remedied when Daya Bay's flux normalization analysis is complete.


\section{Theoretical models of the anomaly}
\subsection{The Sterile Neutrino Solution}
The reactor and gallium anomalies are disappearance anomalies.  Less electron antineutrinos are observed than are predicted by theoretical models.  One consistent explanation is that the models are wrong.   In particular, the theoretical calculations generally have uncertainties of order 2.5\%, and so even if a deficit of 5\% is confirmed the tension with the theoretical calculation will only be 2$\sigma$.  In the remainder of this note we will consider the other possibility, that the anomalies indicate new physics.  The central question is then, what kind of interaction may be responsible for the disappearance?

The simplest explanation would be an interaction.  Perhaps the antineutrinos are simply absorbed as they travel, into the vacuum or the Earth or a dark matter or dark energy field.  In this case one would expect the neutrino density to fall exponentially with the baseline.  This is not observed.  The anomaly at the 100 kilometer distance scales probed by KamLAND is too small to be measured, while it is been observed at the 5\% level at numerous experiments with baselines below 100 meters \cite{reattoreanom}.  Therefore the disappearance cannot be caused by simple absorption.  The anomaly appears to saturate at a short distance.

How can the anomaly saturate?  How can the neutrinos know, after traveling 20 meters, that they have already been absorbed sufficiently and now they should no longer be absorbed?  The only apparent causal explanation is that the neutrino beam travels coherently with another beam, which contains the information concerning how many neutrinos have disappeared.  The simplest realization of this idea is that the electron neutrinos oscillate into another kind of neutrino which then oscillates back, and so the neutrino density reaches  when the ratio of these two kinds of neutrinos reaches a critical value.  Either the new neutrino mass is above the electroweak scale or else LEP can exclude that it interacts weakly.  Strong and electromagnetic interactions for the new neutrino are already excluded by the fact that it can travel through the Earth as far as, say the KamLand experiment.  Therefore the new neutrino is sterile with respect to standard model gauge interactions, although of course dark force interactions are not ruled out.

A new kind of neutrino may sound like a big assumption.  However, if any fermion is neutral under the standard model gauge symmetries then Yukawa interactions with neutrinos and for example the Higgs field $h$ are marginal and are not forbidden and so are expected.  Once such an interaction exists, whatever the nature of the fermion, physicists will call it a sterile neutrino.  This is just one possible coupling, that corresponding to the lowest dimensional operator.  To distinguish it from other couplings, one needs an energy dependence analysis of the anomaly.  Such an analysis is just barely beyond the reach of current Daya Bay data.  As soon as they understand their nonlinear detector response away from 2.5 MeV and have accumulated more data, it may be tested.

What properties must the sterile neutrino have?  To explain the anomaly is quite easy.  To have the right percentage of disappearance one need only fix the mixing angle to be of order $5$ degrees.  The shortest distance reactor and gallium anomalies indicate that the disappearance has saturated by the time the neutrinos have traveled 10 meters, indicating a sterile neutrino mass above 1 eV.  This is all that is needed to be consistent simultaneously with the reactor and gallium anomalies at short baselines and also with limits on neutrino disappearance at longer baselines.  If one allows mixing between sterile and muon neutrinos then this can also explain the anomalous $1-2$ oscillations at LSND \cite{lsnd} and MiniBooNE \cite{minibooneanom,miniboonetuttobene}, although some tension remains with MiniBooNE's lowest energy bins.

\subsection{Cosmological Constraints}

The large sterile neutrino mixing angle required by these anomalies implies that sterile neutrinos will be in thermal equilibrium in the early universe (for recent analyses see \cite{tredueterm,neutrinoasym}).  After they decouple, the sterile neutrino temperature and therefore energy density may be calculated as a function of time.  The energy density impedes the formation of large scale structure (LSS) at scales below the neutrino free streaming length.  This can be compensated by increasing the matter fraction $\Omega_m$.  Type Ia supernova data then prefer a lower value of the dark energy equation of state $w$ whose extra acceleration compensates for the deceleration caused by the additional matter.

However a larger $\Omega_m$ means that there was less time $t_r$ before recombination.    To see this, use the fact that since then the Universe has been matter dominated together with the Friedmann equation to find $t_r(\Omega_m)$
\beq
t_r\sim\frac{2}{3 H(t_r)}\sim\frac{2}{3H_0\sqrt{\Omega_m a(t_r)^{-3}}}.
\eeq
As $a(t_r)$ is fixed by a simple thermodynamic argument together with the CMB temperature, one sees that as $\Omega_m$ increases, $t_r$ decreases.  As the primordial plasma lasted for less time $t_r$, its perturbations do not travel as far before recombination, and so the  baryon acoustic oscillation (BAO) scale is smaller than it would have been with massless neutrinos.

This change in absolute size of the BAO scale is not in itself in contradiction with observations as it is not the absolute scale which is measured, but rather the angular scale and redshift depth.   The measured scale can appear larger, to preserve agreement with observations of the angular size of the BAO peak \cite{baoscoperta,bao}, if the observed galaxies are closer.   As the redshifts of these galaxies are known, they will be closer if the function $z(t)$ is increased.  Equivalently the BAO angular scale can be recoved if $dz/dt$ is increased, which at each redshift is equivalent to increasing the Hubble scale
\beq
H(t)\sim H_0\sqrt{\Omega_m a(t)^{-3}+\Omega_\Lambda a(t)^{-1-3w}}. \label{t2}
\eeq
As $a(t)<1$ it is clear that if $w$ increases then $H$ increases, decreasing the time that light has traveled from these galaxies.  This results in an increased angular size of BAO features.  Thus a larger value of $w$ makes a smaller BAO scale appear bigger and so makes a large value of $\Omega_m$ compatible with BAO observations.  In summary, while reconciling massive neutrinos with supernova requires a smaller value of $w$, baryon acoustic oscillations require a larger value.    This tension provides the strongest cosmological bound on neutrino masses.

Some of this tension is relieved by the presence of an additional flavor of relativistic neutrino at high redshift, increasing the number of effective flavors $N_{\rm{eff}}$ to about 4.   Even if the extra flavor is massless, the preferred dark matter density will increase by about 17\%  \cite{sterilecosm} in order to preserve the redshift at matter-radiation equality
\beq
z_{eq}=-1+\frac{\Omega_m}{\Omega_\gamma(1+0.227 N_{\rm{eff}})} \label{zeq}
\eeq
which is well determined by the CMB power spectrum.  Increasing $N_{\rm{eff}}$ from 3 to 4 increases the denominator in (\ref{zeq}) by about 12\%, compensating for most of the increase in the numerator.  Now when the sterile neutrino masses are turned on, the additional dark matter already somewhat compensates for the free-streaming suppression of medium scale structure.

\subsection{Satisfying the cosmological constraints}

We employ a modified version of CosmoMC \cite{cosmomc} to constrain the sterile neutrino mass by performing a Markov Chain Monte Carlo global fitting analysis with the current observations, which includes the 7-year WMAP temperature and polarization power spectrum \cite{wmap},
the Union2.1 sample of type Ia supernova including systematic errors \cite{Suzuki:2011hu},
BAO distance ratios from SDSS DR7 galaxies \cite{bao}, and $H_0$ \cite{h} as recently measured by the Hubble space telescope.
Since the neutrino masses and the dark energy equation of state lead to correlated effects, we have compared the sterile neutrino mass limits obtained within different dark energy models: the $\Lambda$CDM model and a model with a time-dependent dark energy equation of state
\beq
w(a)=w_0 + w_a(1-a).
\eeq
In the dynamical dark energy models we eliminate the divergences in dark energy perturbations that occur when the dark energy equation of state $w$ crosses -1 using the algorithm suggested in Refs.~\cite{nessundivergenza}.

\begin{figure}[!htb]
\begin{center}
\includegraphics[width=0.6\columnwidth]{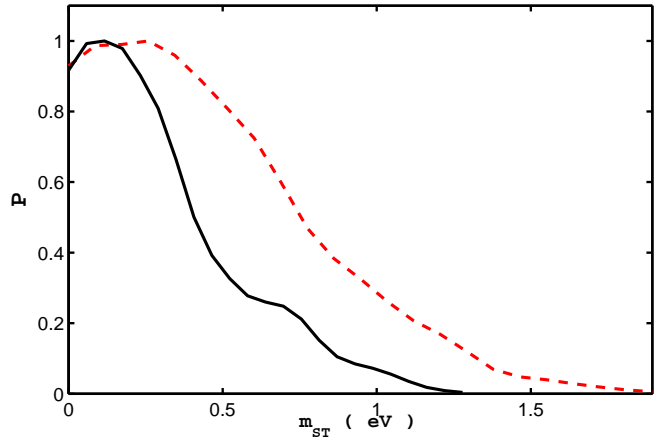}
\caption{1-dimensional frequentist probability distribution of the sterile neutrino mass given by fitting with WMAP7, BAO and the Union2.1 supernova dataset. The black solid line is given by fitting with $\Lambda$CDM model, while the red dashed line is obtained using a time evolving dark energy model}.
\label{mfit}
\end{center}
\end{figure}

First, we determined the compatibility of neutrinos of the desired mass with our cosmological model.  While the reactor anomaly can be caused by sterile neutrinos as light as $0.5$ eV, such neutrinos would be irrelevant at the short distances probed by the gallium anomaly.  Similarly $1$ eV neutrinos yield a somewhat better fit for the LSND and MiniBOONE anomalies than $0.5$ eV neutrinos, considering the bounds on the mixing angles from other experiments.
Assuming that there are 3 massless neutrinos and 1 (sterile) massive neutrino, we confront the model with current data, and in Fig. {\ref{mfit}} we plot the $1$-dimensional frequentist probability distribution of the sterile neutrino mass.  By fitting with different dark energy parametrizations, we obtain $2\sigma$ constraints on the sterile neutrino mass of $m_{ST} < 0.8~eV$, and $m_{ST}<1.1~ eV$ respectively for $\Lambda$CDM and for time-evolving dark energy models.  In the $\Lambda$CDM model, the $0.5$ eV sterile neutrino is compatible within $2\sigma$ and the $1$ eV model, preferred by neutrino experiments, is excluded by more than $2\sigma$, while the tension between these observations and 1 eV massive sterile neutrinos can be reduced within the dynamical dark energy models.

To better understand the cosmological consequences of massive sterile neutrinos, we then restricted our attention to 3 flavors of light active neutrinos and a single 1 eV sterile neutrino.  We found that cosmology prefers a larger dark matter density  $\omega_{\rm{cdm}}=\Omega_{\rm{cdm}}h^2$ and a lower value of $w_a$ than in the massless case with no sterile neutrino.   Our mean preferred values and $1\sigma$ constraints are $w_a=-1.681\pm1.136,~\omega_{\rm{cdm}}=0.138\pm0.005$, which can be compared with the no sterile neutrino case $w_a=-0.862\pm1.009,~\omega_{\rm{cdm}}=0.116\pm0.005$.

\begin{figure}[!htb]
\begin{center}
\includegraphics[width=0.6\columnwidth]{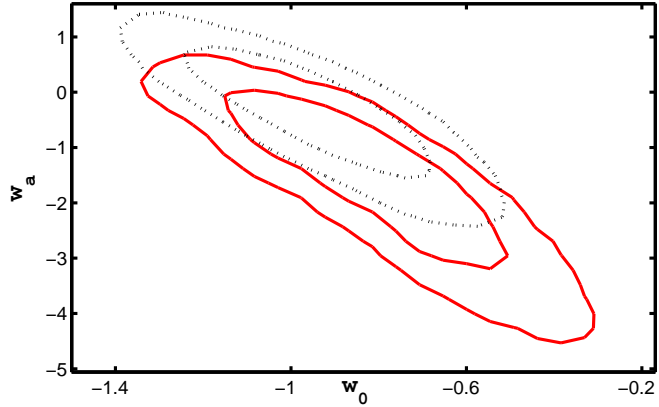}
\caption{2-dimensional cross correlation constraints on $w_0$ and $w_a$ obtained by fitting with WMAP7, BAO and the Union2.1 supernova dataset. The red solid lines and black dot lines are the 1 and 2$\sigma$ compatibility regions with and without a single $1~eV$ sterile neutrino flavor respectively. }.
\label{fpEOS}
\end{center}
\end{figure}

In Fig.~\ref{fpEOS} we plot the cross correlation contours of $w_0$ and $w_a$ with and without $1~eV$ sterile neutrinos. The red solid and the black dotted curves bound the 1 and 2$\sigma$-allowed regions with and without the sterile neutrinos respectively.  The red contours are shifted towards a higher $w_0$ and a more negative value of $w_a$ as compared with the black contours.  This means that a higher value of $w$ is preferred at low redshifts, where BAO data dominates, and a lower value at redshifts beyond the reach the SDSS large scale structure surveys, where supernova data dominates.  The redshift dependence that we found is much stronger than that found in Ref.~\cite{altroquintom}, reflecting the fact that the Union2.1 supernova dataset that we used contains about twice as many supernova at redshift $z>1$ than the Union2 dataset \cite{unione2} used in their analysis.

Note that our solution to the cosmological constraints is deep inside of the quintom regime \cite{quintom}, as $w_a$ is far from zero.  It may be that it only provides a fair fit to the data because BAO constraints are weak at high redshifts, beyond the distances at which luminous red galaxies have been surveyed.  However the steeply sloped dark energy equation of state in such models strongly favor a very large BAO acoustic scale at high redshifts, a prediction which will soon be tested.

As one might expect, more radical solutions can be yet more compatible with the cosmological data, such as a high sterile neutrino initial asymmetry \cite{neutrinoasym} which can prevent sterile neutrinos from thermalizing \cite{neutrinoasymvecchio} or a modified theory of gravity \cite{fr}.   The suppression of small scale structure formation can be compensated by including extended objects in the early universe, such as cosmic strings \cite{robert} or giant monopoles \cite{monopoli}, obviating the need for an increased dark matter fraction.  Nonstandard matter effects may also be responsible for these anomalies, for example in Ref.~\cite{lsndnonstandard} the authors used sterile neutrinos with masses within cosmological bounds subjected to nonstandard matter effects to explain the LSND and MiniBooNE anomalies.

The tension may also be eliminated altogether if the sterile neutrino couplings are environmentally-dependent \cite{envir}, for example in a chameleon model in which the sterile neutrino mass is proportional to the density of its environment.  It was argued in Ref.~\cite{neal04} that, in the case of the couplings here, such models are consistent with both precision gravity and equivalence principle bounds.   LSND, MiniBooNE, reactor anomaly and gallium anomaly experiments all have a significant fraction of the neutrino baseline inside of dense media like the Earth and detector shielding.  If the sterile neutrino mass is proportional to the background density and  the sterile neutrino mass is of order 2 eV inside of the Earth then in space it will be relativistic at all times since decoupling from the primordial plasma.  Thus, so far as big bang nucleosynthesis, structure formation and CMB and baryon oscillations are concerned, the sterile neutrino is massless and all cosmological bounds are easily satisfied.  In fact, the addition of a massless neutrino in general improves the cosmological fits compared with with the $\Lambda$CDM model.

The fact that about half of the LSND and MiniBooNE baseline is air means that the required neutrino mass will increase to about 2 eV, which is still compatible with constraints from beta decay.  However, as the KARMEN baseline has a higher fraction of air, the historical tension between LSND and KARMEN will be reduced.  This dependence upon the material in the baseline may be quite easy to test.  A minimum amount of shielding is always required around a detector, however this can be thinner in the path followed by the neutrinos.  Of course, nuclear power plants themselves always have shielding, and it may be that this shielding is already sufficient for the anomalous oscillation to occur.  In this case, gallium experiments may be a more promising setting to test the environmental dependence of sterile neutrino masses.

\section{Conclusions}
Combined with cosmological evidence for an additional relativistic degree of freedom before recombination, the very short distance reactor, gallium and LSND/MiniBooNE neutrino disappearance anomalies provide an ever more substantial argument for a heavy sterile neutrino.  While neutrino disappearance has also been observed at baselines ranging from 100 m to 2 km, it has not been so far possible to determine how much of this disappearance is attributable to 1-3 neutrino flavor oscillations and so it has not been possible to use these experiments to support or refute the reactor anomaly hypothesis.

With the flux normalization independent precision measurements of the 1-3 mixing angle at Daya Bay and RENO the degeneracy between the reactor anomaly and $\theta_{13}$ has been broken.   Therefore, although these experiments provide only moderately precise measurements of the flux normalization themselves, their determinations of $\theta_{13}$ may be applied to all neutrino oscillation experiments with baselines longer than 100 meters to determine just how much 1-3 oscillation has occurred, and so what remaining signal is anomalous.  This could allow a more precise reanalysis at accelerator experiments such as T2K and MINOS.  In this note we applied them to the simpler setting of the short baselines reactor experiments themselves.  By combining the disappearance results of Double Chooz, Daya Bay and RENO we were able to simultaneously fit $\theta_{13}$ and the anomalous disappearance.  We found that these experiments certainly do not exclude a reactor anomaly and indeed lend it some additional support.

Our results are still strongly dependent upon assumptions concerning the theoretic flux at RENO and are bounded by the statistical error at Daya Bay.  However both of these shortcomings will remedied by data soon to be released by these experiments, and then we expect that an application of such 2-dimensional fits to the new data will provide a somewhat more precise determination of the reactor anomaly, complimenting the scheduled very short distance measurements.

The theoretical errors in the normalized reactor fluxes remain quite large.  And so even if the existence of a reactor anomaly may be convincingly demonstrated in the near future, it will be much more difficult to determine whether it is attributable entirely to error in this calculation or else to new physics.  Indeed it is quite plausible that a convincing determination will require inputs from cosmology, which are only weakly dependent upon mixing angles and mass differences but are rapidly providing a more convincing case for a sterile neutrino and potentially, in a few years, a value for at least the total neutrino mass. Therefore we have in this paper also attempted to analyze the consistency of sterile neutrino models with cosmological constraints.

Until last year, CP violating anomalies suggested that there be at least 2 flavors of sterile neutrinos, which were in general strongly disfavored by cosmological constraints.  Evidence for CP violation in neutrino anomalies has since weakened appreciably, and so now models with 1 flavor of sterile neutrino provide reasonable fits to experimental data.  The best fits are obtained for neutrino masses near $1$\ eV.  As we have reviewed, in a $\Lambda CDM$ cosmology such massive sterile neutrinos are essentially excluded, and even allowing a constant dark energy equation of state $w$ they cannot be made simultaneously compatible with BAO and supernova observations.   

However we observe that the various constraints on neutrino masses are relevant at different redshifts.  BAO constraints are tight at low redshifts $z\sim 0.2-0.3$ and supernova constraints are tighter at higher redshifts, in particular using the Union2.1 dataset which includes many more supernova at $z>1$ than its predecessors.  Therefore, at least for now, a strong redshift dependence in $w$ significantly reduces the tension between these cosmological constraints, while making strong predictions for future, higher redshift, large scale structure surveys.  

Further we found that the tension may be eliminated entirely by noting that all evidence for massive neutrinos occurs inside of dense media whereas constraints on neutrino masses occur in much lower density environments.  Therefore models in which the sterile neutrino mass depends on the background density are consistent with both.  These models have the additional attractive feature that they explain the observed anomalous oscillation at LSND despite the negative results at KARMEN.  While KARMEN's total baseline is more than half of that of LSND, LSND has appreciably more dense shielding.  The density weighted baseline at LSND is appreciably higher, increasing the amount of oscillation expected at LSND with respect to KARMEN.  More importantly, such models are easy to test, in particular at the next generation of small experiments designed to test the reactor anomaly.  Changing the ratio of air to shielding between the reactors and detectors when possible, or otherwise changing the density of the shielding will in general affect the neutrino oscillation probability in such models, allowing them to be easily excluded in the near future.

\section* {Acknowledgement}

\noindent
We would like to thank Xinmin Zhang for useful discussions and advice in every phase of this project. We are also grateful to Thierry Lasserre and S\"oren Jetter for patient discussions and correspondence.  We acknowledge the use of the Legacy Archive for Microwave
Background Data Analysis (LAMBDA). Support for LAMBDA is provided by
the NASA Office of Space Science. The calculation was performed at the
Deepcomp7000 of Supercomputing Center, Computer Network Information
Center of Chinese Academy of Sciences.
JE is supported by the Chinese Academy of Sciences
Fellowship for Young International Scientists grant number
2010Y2JA01. EC is supported in part by the NSF of
China.  HL is supported by the National Science Foundation
of China under Grant Nos.~11033005, the 973 program
under Grant No.~2010CB83300 and the Chinese Academy
of Science under Grant No. KJCX2-EW-W01.


\end{document}